\documentclass[11pt]{article}

\usepackage{stfloats}

\usepackage{fancyhdr}
\usepackage{balance}
\usepackage{pstricks}
\usepackage{import}
\usepackage{cite}
\usepackage{graphicx}
\usepackage{acronym}
\usepackage{amsmath} 
\usepackage{subfigure}
\newcommand{\comments}[1]{} 
\usepackage{float}
\usepackage{amsthm}
\usepackage{amssymb} 
\usepackage{algorithmic}
\usepackage{algorithm}
\usepackage{balance}
\DeclareMathOperator*{\argmin}{arg\,min}
\DeclareMathOperator*{\argmax}{arg\,max}

\newcommand{\ie}{\emph{i.e.}} 
\newcommand{\eg}{\emph{e.g.}}

\newcommand{\B}{\boldsymbol}

\acrodef{OPEX}[OPEX]{Operating Expenses}
\acrodef{UE}[UE]{User Equipment}
\acrodef{BS}[BS]{Base Station}
\acrodef{DTX}[DTX]{Discontinuous Transmission}
\acrodef{PAPR}[PAPR]{Peak-to-Average Power Ratio }
\acrodef{SC-FDMA}[SC-FDMA]{Single-carrier FDMA}
\acrodef{FDMA}[FDMA]{Frequency Division Multiple Access}
\acrodef{TDMA}[TDMA]{Time Division Multiple Access}
\acrodef{CDMA}[CDMA]{Code Division Multiple Access}
\acrodef{OFDMA}[OFDMA]{Orthogonal Frequency Division Multiple Access}
\acrodef{OFDM}[OFDM]{Orthogonal Frequency Division Multiplexing}
\acrodef{ICT}[ICT]{Information and Communication Technologies}
\acrodef{QoS}[QoS]{Quality of Service}
\acrodef{PA}[PA]{Power Amplifier}
\acrodef{RS}[RS]{Resource Sharing}
\acrodef{PC}[PC]{Power Control}
\acrodef{SOTA}[SotA]{State-Of-The-Art}
\acrodef{EE}[EE]{Energy Efficiency}
\acrodef{SINR}[SINR]{Signal-to-Interference-and-Noise-Ratio}
\acrodef{LTE}[LTE]{Long Term Evolution}
\acrodef{EARTH}[EARTH]{Energy Aware Radio and neTwork tecHnologies}
\acrodef{MIMO}[MIMO]{Multiple-Input Multiple-Output}
\acrodef{SISO}[SISO]{Single-Input Single-Output}
\acrodef{SIMO}[SIMO]{Single-Input Multiple-Output}
\acrodef{RE}[RE]{Resource Element}
\acrodef{SNR}[SNR]{Signal-to-Noise-Ratio}
\acrodef{ACLR}[ACLR]{Adjacent Carrier Leakage Ratio}
\acrodef{BA}[BA]{Bandwidth Adaptation}
\acrodef{MCS}[MCS]{Modulation and Coding Scheme}
\acrodef{AA}[AA]{Antenna Adaptation}
\acrodef{RRM}[RRM]{Radio Resource Management}
\acrodef{WSN}[WSN]{Wireless Sensor Network}
\acrodef{CSI}[CSI]{Channel State Information}
\acrodef{RRH}[RRH]{Remote Radio Head}
\acrodef{RCG}[RCG]{Rate Craving Greedy}
\acrodef{RF}[RF]{Radio Frequency}
\acrodef{3GPP}[3GPP]{3rd Generation Partnership Project}
\acrodef{RAPS}[RAPS]{Resource allocation using Antenna adaptation, Power control and Sleep modes}

\begin{document}
%
\title{Minimizing Base Station Power Consumption}
%
%
%

\author{{Hauke~Holtkamp, Gunther~Auer, Samer Bazzi}\vspace{3mm}\\
DOCOMO Euro-Labs\\
D-80687 Munich, Germany \\Email:
\{lastname\}@docomolab-euro.com
\and Harald~Haas\vspace{2mm}\\
Institute for Digital Communications\\
Joint Research Institute for Signal and Image Processing\\
 The University of Edinburgh,
EH9 3JL, Edinburgh, UK\\ E-mail: h.haas@ed.ac.uk}%
\markboth{Journal of Selected Areas in Communications,~Vol.~31, No.~5, May~2013}%
{Minimizing Base Station Power Consumption}


\maketitle

\begin{abstract}
We propose a new radio resource management algorithm which aims at minimizing the base station supply power consumption for multi-user MIMO-OFDM. Given a base station power model, that establishes a relation between the RF transmit power and the supply power consumption, the algorithm optimizes the trade-off between three basic power-saving mechanisms: antenna adaptation, power control and discontinuous transmission. The algorithm comprises two steps: a) the first step estimates sleep mode duration, resource shares and antenna configuration based on average channel conditions and b) the second step exploits instantaneous channel knowledge at the transmitter for frequency selective time-variant channels. The proposed algorithm finds the number of transmit antennas, the RF transmission power per resource unit and spatial channel, the number of discontinuous transmission time slots, and the multi-user resource allocation, such that supply power consumption is minimized. Simulation results indicate that the proposed algorithm is capable of reducing the supply power consumption by between 25\% and 40\%, dependend on the system load. 

\end{abstract}


\acresetall

\section{Introduction}
%
%


The power consumption of cellular radio networks is constantly increasing due to the growing number of mobile terminals and higher traffic demands, which require a densification of the network topology. Environmental concerns, rising energy costs and a growing need for self-sufficient (power-grid independent) \acp{BS} reinforce efforts to reduce the cellular networks' power consumption~\cite{fmbf1001}. Shortcomings of the state-of-the-art lie in the fact that current \acp{BS} are designed to serve peak demands without considering energy efficient off-peak operation. Depending on the time of day or geographic location, a \ac{BS} may be idle---and thus over-provisioning---for a significant portion of its operation time. In such low-load situations, spectral efficiency can be traded for energy efficiency without reducing the user experience. The importance here is to operate \acp{BS} such that they can flexibly adjust to traffic demand. Furthermore, any power-saving operation on the \ac{BS} side must not negatively affect the mobile terminal, \eg\ by increasing its power consumption or reducing the perceived quality of service.

With regard to different power-saving \ac{RRM} mechanisms, \ac{PC} is the most prominent in literature. This is due to the fact that, in addition to reducing power consumption, \ac{PC} is also beneficial to link adaptation and interference reduction~\cite{wclm9901,cgb0401}. In \cite{wclm9901} Wong \emph{et al.}\ minimize transmit power for multi-user \ac{OFDM} under rate constraints. Their \ac{PC} algorithm is computationally complex and does not consider a transmit power constraint or power model. Cui \emph{et al.} analyze the energy efficiency of \ac{MIMO} transmissions, being the first to consider the supply power consumption in energy efficiency studies~\cite{cgb0401}. They show that in terms of energy efficiency, the optimal number of transmit antennas used for \ac{MIMO} transmission depends on the \ac{SNR}. Miao \emph{et al.}~\cite{mhlb0801} derive the data rate that maximizes the transmitted information per unit energy (bit per joule). More recently, acknowledging the significance of circuit power consumption, \ac{DTX} and \ac{AA} have been identified as energy saving \ac{RRM} techniques. Kim \emph{et al.}~\cite{kcvh0901} establish \ac{AA} as a \ac{MIMO} resource allocation problem and adapt the number of transmit antennas on a single link. However, solutions provided for \ac{AA} on the mobile side cannot be directly applied to the \ac{BS}, since for the latter multiple links have to be considered. Some works have assessed the energy saving potentials of sleep modes and \ac{AA} in \ac{LTE} \acp{BS}, but have applied very crude decision mechanisms~\cite{hafm1201,fmmjg1101,wthg1101}, \eg\ a \ac{BS} is only allowed to sleep when a cell is empty. To the best of our knowledge, there are no prior works which consider \ac{DTX} in combination with other power saving \ac{RRM} mechanism. 

The closest related works to Inverse Water-filling, which is introduced in Section~\ref{step1}, have studied the problem of allocating transmit power, user rates and resources per user with suboptimal solutions~\cite{cgb0401,sae0301,aw1001}. Bit capacities, modulation as well as transmission power consumption are investigated in general \ac{MIMO} point-to-point transmissions~\cite{cgb0401}, but without considering power-allocation to orthogonal channels. A later work assigns power sub-optimally in favor of a fairness constraint~\cite{sae0301}. Al-Shatri \emph{et al.}~\cite{aw1001} have used Lagrange multipliers to invert water-filling in a rate-maximizing fairness application. We extend their work by relaxing the equal rate assumption among users and present a complete power-minimizing algorithm.

In this paper, we address the issue of overly generic assumptions about underlying hardware by identifying the relevant energy saving schemes for cellular networks on the basis of a realistic and detailed power model~\cite{aggsoisgd1101}. As a consequence, and unlike previous energy saving studies~\cite{wclm9901}, this work minimizes the \textit{supply power} (or 'AC-plug' power) of the cellular \ac{BS} rather than the \ac{RF} transmission power. To provide an applicable mechanism for current cellular systems like \ac{3GPP} \ac{LTE}, we present the \ac{RAPS} algorithm which reduces the \ac{BS} supply power consumption of multi-user \ac{MIMO}-\ac{OFDM}. Given the channel states and target rates per user, \ac{RAPS} finds the number of transmit antennas, the number of \ac{DTX} time slots and the resource and power allocation per user. \ac{DTX} refers to a micro sleep on the transmitter side, which is sufficiently short so that the regular operation of the mobile is not affected. The \ac{RAPS} solution is found in two steps: First, \ac{PC}, \ac{DTX} and resource allocation are joined into a convex optimization problem which can be efficiently solved. Second, subcarrier and power allocation for a frequency-selective time-variant channel pose a combinatorial problem, which is solved by means of a heuristic solution. To allocate minimal power levels to resources and spatial channels, we derive the Inverse Water-filling algorithm.

The paper is organized as follows: The system and power model is presented in Section~\ref{model}. The global energy efficiency problem is defined in Section~\ref{problemstatement}. Section~\ref{step1} presents Step~1 of the two-step \ac{RAPS} algorithm which determines \ac{DTX} duration, antenna number, and resource shares on the basis of block fading channels. Step~2 extends these estimates by subcarrier and power allocation in frequency-selective time-varying channels in Section~\ref{step2}. Simulation results are presented in Section~\ref{results}. The paper is concluded in Section~\ref{conclusion}.
%

\section{System and power model}
\label{model}

\subsection{System Model}
\begin{figure}
\centering
\includegraphics[width=0.6\textwidth]{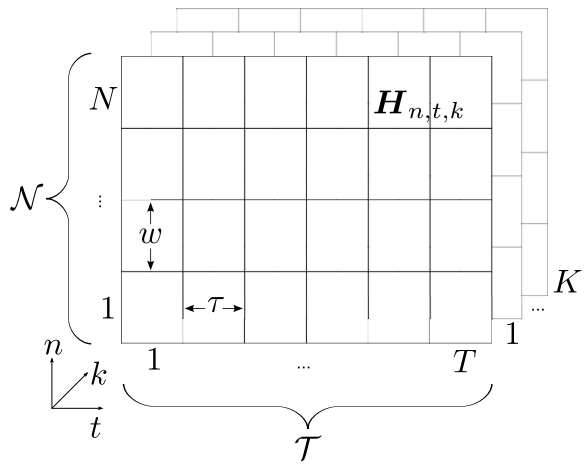}
\caption{OFDM frame structure.}
\label{fig:framestructure}
\end{figure}

We consider one transmission frame in the downlink of a point-to-multipoint wireless communication system, comprising one serving \ac{BS} and multiple mobile receivers. The \ac{BS} transmitter is equipped with $M_{\mathrm{T}}$ antennas and all antennas share the transmit power budget. Mobile receivers have $M_{\mathrm{R}}$ antennas and system resources are shared via \ac{OFDMA} between $K$ users on $N$ subcarriers and $T$ time slots. In total there are $T N$ resources units. \ac{OFDMA} is used, for example, in \ac{3GPP} \ac{LTE}. A frequency-selective time-variant channel is assumed, with each resource unit characterized by the channel state matrix $\boldsymbol{H}_{{n,t,k}} \in \mathbb{C}^{M_{\mathrm{R}} \times M_{\mathrm{T}}}$, with subcarrier index $n = 1,\ldots,N$, time slot index $t = 1,\ldots,T$, user index $k = 1,\ldots,K$. 
The vector of spatial channel eigenvalues per resource unit~$a$ and user~$k$ is $\mathcal{E}_{a,k}$ and its cardinality is $\min\{M_{\mathrm{T}},M_{\mathrm{R}}\}$. The system operates orthogonally such that individual resource units cannot be shared among users. \ac{MIMO} transmission with a variable number of spatial streams is assumed over the set of resources assigned to each user $\mathcal{A}_k$. The frame structure is illustrated in Fig.~\ref{fig:framestructure}. 


Although cellular wireless networks are generally interference limited, co-channel interference between neighbouring cells is not considered in this work. The main contribution of this work lies in the detailed derivation of the \ac{BS} supply power minimization via a power model. The consideration of interference and resulting network dynamics are, therefore, beyond scope of this paper and are left for future work.

\subsection{Power Model}
Supply power consumption, \ie\ the overall device power consumption of the \ac{BS}, is calculated via a power model. Power consumption models for cellular \acp{BS} are established in~\cite{aggsoisgd1101,aggsoigdb1101}. The model allows to map \ac{RF} transmission power to supply power consumption. The power model is constructed from a real \ac{BS} hardware implementation, and is able to capture the most relevant power consumption effects in commercially available \acp{BS}. It provides a simple linear model which is verified by an advanced realistic and detailed hardware model~\cite{ddgfahwsr1201}. Therefore, the linear power model provides a sufficient foundation for the \ac{BS} energy efficiency analysis conducted in this paper. In the power model~\cite{aggsoisgd1101,aggsoigdb1101}, a \ac{RF} chain refers to a set of hardware components composed of a small-signal transceiver section, a \ac{PA} and the antenna interface. The \ac{PA} consumes a large share of the overall power due to its low efficiency~\cite{aggsoisgd1101}. Each transmit antenna is connected to an \ac{RF} chain, thus the term \ac{AA} implies the adaptation of the respective \ac{RF} chain. Unlike, \eg\ a shared baseband component, it is assumed that \ac{RF} chains can be switched off or put to sleep individually when there is no transmission on the respective antenna. The \ac{BS} consumes less power when fewer \ac{RF} chains are active. Setting the entire \ac{BS} to sleep interrupts transmission but further reduces the power consumption. For simplicity, we assume instantaneous and cost-free on-off-switching of hardware, \ie\ there is no delay or additional power cost incurred by putting a \ac{BS} into \ac{DTX} mode or enabling/disabling an \ac{RF} chain. This assumption is supported by~\cite{fmmjg1101}, in which a switching time of 30\,$\rm{\mu}$s is assumed, which is much smaller than the duration of an \ac{OFDMA} frame. We, therefore, presume that the additional power penalty due to the settling time of the \ac{RF} components can be neglected.

\begin{figure}
\centering
 \includegraphics[width=0.6\textwidth]{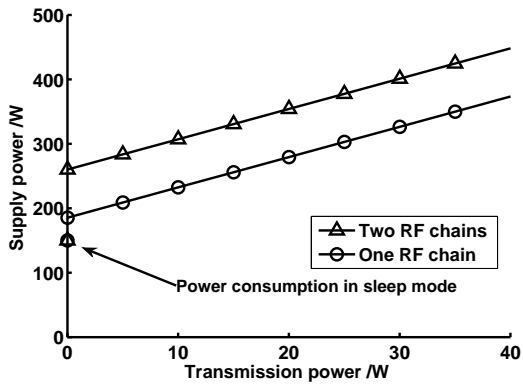}
\caption{Supply power model~\cite{aggsoisgd1101}.}
\label{fig:powermodel}
\end{figure}

Power model parameters are provided in~\cite{aggsoisgd1101} only for a \ac{BS} with two \ac{RF} chains. We derive the supply power consumption value when operating with a single \ac{RF} chain as follows: In~\cite{aggsoisgd1101} power consumption characteristics for a macro \ac{BS} with three sectors are provided. The power consumption is rated at a minimum active consumption $P_{0,M_{\mathrm{T}}}$, which depends on $M_{\mathrm{T}}$, a linear transmission power dependence factor $\Delta_{\mathrm{PM}}$ (the slope) and a power consumption in \ac{DTX} mode $P_{\mathrm{S}}$. Transmission power consumption in any time slot~$t$ is $P_t$. Under the assumption of unchanged efficiencies of all components, we use the tables provided in \cite{aggsoisgd1101} to find the maximum power consumption of a \ac{BS} operated with a single \ac{RF} chain and assume $\Delta_{\mathrm{PM}}$ to be unchanged. Since we presume instantaneous switching capabilities for all components, the sleep mode consumption is the same in both \ac{RF} chain settings. Lastly, we normalize the power consumption values to a single sector, leading to the power model depicted in Fig.~\ref{fig:powermodel}. Formally, the supply power consumption is
\begin{equation}
\label{eq:psupply}
 P_{\rm{supply}}(P_t) = 
  \begin{cases}
   P_{0,M_{\mathrm{T}}} + \Delta_{\mathrm{PM}} P_t, 			& \text{if } P_t > 0, \\
   P_{\mathrm{S}},			      	& \text{if } P_t = 0.
  \end{cases}
\end{equation}

\section{Global Problem Statement}
\label{problemstatement}

In this section, we deduce power-saving mechanisms from the power model and derive the global \ac{OFDMA} supply power minimization problem. 

\subsection{Energy Saving Strategies}
The power model in Fig.~\ref{fig:powermodel} reflects three general mechanisms that affect the power consumption of \acp{BS}: First, the overall power emitted at the \ac{PA} $P_t$ affinely relates to supply power consumption~\eqref{eq:psupply}. Second, if the same \ac{BS} is operated with fewer \ac{RF} chains, it consumes less power. Third, if the \ac{BS} is put into sleep mode, it consumes less power than in the active state with zero transmission power.

These observations lead to the following saving strategies: 
\begin{description}
 \item[\ac{PC}]: Reduce transmission power on each resource unit.
 \item[\ac{AA}]: Reduce the number of \ac{RF} chains.
 \item[\ac{DTX}]: Increase the time the \ac{BS} spends in \ac{DTX}.
\end{description}
The first and the last strategy are clearly opposing each other as lower transmission powers lead to lower link rates and thus longer transmission duration (for an equal bit-load), whereas it would be beneficial for long \ac{DTX} to have short transmissions. The second and third strategy are related, as \ac{AA} can be considered a weak form of \ac{DTX} which still allows transmission on a subset of antenna elements. 

\subsection{Joint Strategies}

\begin{figure}
 \includegraphics[width=0.6\textwidth]{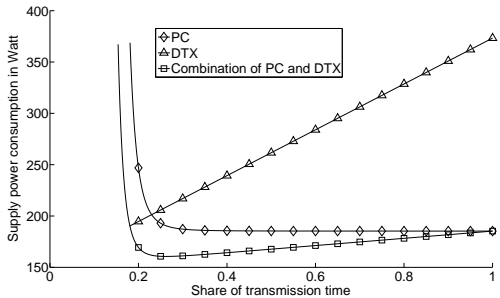}
\caption{Supply power consumption for transmission of a target spectral efficiency as a function of time spent transmitting, $\Phi$. The combination of \ac{DTX} and power control achieves lower power consumption than exclusive operation of either DTX or power control.}
\label{fig:PCsleeptradeoff}
\end{figure}

For a single link, the energy efficiency trade-off between \ac{PC} and \ac{DTX} is illustrated in Fig.~\ref{fig:PCsleeptradeoff}. Shown is the supply power consumption caused by transmitting a fixed amount of data. 
Three operation modes are depicted in Fig.~\ref{fig:PCsleeptradeoff}:
\begin{itemize}
 \item \ac{PC}: Only \ac{PC} but no \ac{DTX} is available. The transmission power is adjusted depending on the transmission time normalized to the time slot duration, $\Phi$, such that the target bit load is transmitted. The \ac{BS} consumes idle power $P_0$ when there is no data transmission. Clearly, the strategy for lowest power consumption in this case is to set $\Phi{=}1$. Reducing $\Phi$ increases the required transmission power, until at $\Phi=0.18$, $P_{\mathrm{Tx}} = P_{\mathrm{max}}$.
\item \ac{DTX}: The \ac{BS} always transmits with full power $P_{\mathrm{Tx}} = P_{\mathrm{max}}$. The supply power is $P_0 + m P_{\mathrm{max}}$ when transmitting or $P_{\mathrm{S}}$ when in \ac{DTX} mode, yielding an affine function of $\Phi$. At $\Phi=1$, more data than the target bit load is transmitted. Reducing $\Phi$ from $\Phi=1$ decreases the supply power consumption linearly up to the point where the target bit load is met. Here, the best strategy clearly is to minimize the time transmitting (small $\Phi$), so to maximize the time in \ac{DTX} ($1{-}\Phi$). 
\item Joint application of \ac{PC} and \ac{DTX}: In this mode of operation, the \ac{DTX} time ($1{-}\Phi$) is gradually reduced. The transmission power is adjusted to meet the target bit load within~$\Phi$. 
\end{itemize}

Fig.~\ref{fig:PCsleeptradeoff} shows that the joint operation of \ac{PC} and \ac{DTX} consumes less power than each individual mode of operation with an optimal point at $\Phi{=}0.25$. 

\subsection{Global Problem}

We extend the power control and \ac{DTX} trade-off to \ac{MIMO}-\ac{OFDM} serving multiple users over frequency-selective channels. The selection of the number of transmit antennas for \ac{AA} is made once for the entire frame. 

We formulate the global problem statement: Given the channel state matrices $\B{H}_{n,t,k}$ on each channel and the vector of target rates per user $\B{r}=(R_1,\dots,R_K)$, we seek:
\begin{itemize}
 \item the set of resources allocated to each user $\mathcal{A}_{k}$, 
 \item the power level $P_{a,e}$ per resource unit $a = 1,\dots, |\mathcal{A}_{k}|$ and spatial channel $e =1,\dots,|\mathcal{E}_{a,k}|$,
 \item and the number of active transmit antennas $M_{\mathrm{T}}$, 
\end{itemize}
such that the supply power consumption is minimized while fulfilling the transmission power constraint~$P_{\mathrm{max}}$.

The sum capacity of user~$k$ over one transmission frame of duration~$\tau_{\mathrm{frame}}$ is given by:
\begin{equation}
\label{eq:R_k}
 R_k = \frac{w \tau}{\tau_{\mathrm{frame}}} \displaystyle\sum_{a=1}^{|\mathcal{A}_{k}|} \sum_{e = 1}^{|\mathcal{E}_{a,k}|} \log_2 \left( 1 + \frac{P_{a,e} \mathcal{E}_{a,k}(e)}{N_0 w} \right),
\end{equation} 
with subcarrier bandwidth $w$ in Hz, time slot duration $\tau$ in seconds, and noise spectral density $N_0$ in W/Hz.

The RF transmission power in time slot~$t$ is
\begin{equation}
\label{eq:Pt}
 P_t = \displaystyle \sum_{a=1}^{|\mathcal{A}_{t}|} \sum_{e=1}^{|\mathcal{E}_{a}|} P_{a,e},
\end{equation}
where $\mathcal{A}_{t}$ is the set of $N$ resources in time slot $t$ and $\mathcal{E}_{a}$ is the vector of channel eigenvalues on resource~$a$.

Given \eqref{eq:psupply}, \eqref{eq:R_k} and \eqref{eq:Pt}, the optimization problem that minimizes the supply power consumption is of the form
\begin{equation}
 \begin{gathered}
  \hspace*{-23mm}\underset{   \mathcal{A}_{k}\forall k, P_{a,e}\forall a,e, M_{\mathrm{T}}    }    {\text{minimize}} \quad P_{\mathrm{supply,frame}} (\B{r}) =		\\
   \frac{1}{T} \left( \displaystyle \sum_{t=1}^{T_{\mathrm{Active}}} \left( P_{0,M_{\mathrm{T}}} + \Delta_{\mathrm{PM}} P_t \right) + \sum_{t=1}^{T_{\mathrm{Sleep}}} P_{\mathrm{S}} \right) \\
  \hspace*{-10mm}\text{subject to} \quad \displaystyle P_t \leq P_{\mathrm{max}}\,, \ 
  T_{\mathrm{Active}} + T_{\mathrm{Sleep}} = T,\\
  \hspace*{5mm} R_k \leq \frac{w \tau}{\tau_{\mathrm{frame}}} \displaystyle\sum_{a=1}^{|\mathcal{A}_k|} \sum_{e=1}^{|\mathcal{E}_{a,k}|} \log_2 \left( 1 + \frac{P_{a,e} \mathcal{E}_{a,k}(e)}{N_0 w} \right)
 \end{gathered}
\end{equation}
with the number of active transmission time slots $T_{\mathrm{Active}}$, and \ac{DTX} slots $T_{\mathrm{Sleep}}$. This is a set selection problem over the sets $\mathcal{A}_k$, and $\mathcal{E}_{a,k}$, as well as a minimization problem in~$P_{a,e}$. 

\subsection{Complexity}
Dynamic subcarrier allocation is known to be a complex problem for a single time slot in frequency-selective fading channels that can only be solved by suboptimal or computationally expensive algorithms~\cite{wclm9901,kll0301,jl0301}. In this study, we add two additional degrees of freedom by considering \ac{AA} and \ac{DTX}, increasing the complexity further. We consequently divide the problem into two steps: First, real-valued estimates of the resource share per user, \ac{DTX} duration, and number of active \ac{RF} chains $M_{\mathrm{T}}$, are derived based on simplified system assumptions. Second, the power-minimizing resource allocation over the integer set $\mathcal{A}_k$ and consecutive Inverse Water-filling power allocation are performed.

%

\section{Step~1: Antenna Adaptation, \ac{DTX} and Resource Allocation}
\label{step1}
The first step to solving the global problem is based on a simplification of the system assumptions. This allows defining a convex subproblem with an optimal solution, which is later (sub-optimally) mapped to the solution of the global problem in Step~2, described in Section~\ref{step2}. 

Instead of time and frequency-selective fading we assume that channel gains on all resources are equal to the center resource unit:
\begin{equation}\label{eq:chnCenter}
 \B{H}_k = \B{H}_{n^{\mathrm{c}}, t^{\mathrm{c}},k},
\end{equation}
where the superscript ${\mathrm{c}}$ signifies the center-most subcarrier and time slot. Step~1 thus assumes a block fading channel per user over $W=N w$ and $\tau_{\mathrm{frame}}$. 

We select the center resource unit due to the highest correlation with all other resources. Alternative methods to construct a representative channel state matrix are to take the mean or median of $\B{H}_{n,t,k}$ over the \ac{OFDMA} frame. However, application of the mean or median over a set of \ac{MIMO} channels were found to result in a channel with lower capacity. 
See Section~\ref{results} for a comparison plot between different channel selection methods.

The link capacity is calculated using equal-power precoding and assuming uncorrelated antennas. In contrast to water-filling precoding, equal-power precoding provides a direct relationship between total transmit power and target rate. On block fading channels with real-valued resource sharing, OFDMA and \ac{TDMA} are equivalent. Without loss of generality, we select resource allocation via \ac{TDMA}. These simplifications allow to establish a convex optimization problem that can be efficiently solved.


In a block fading multi-user downlink with target rate~$R_k$ and equal power precoding, the spectral efficiency for user~$k$ is given by
\begin{equation}
  C_k = \frac{R_k}{W \mu_k} = \sum_{e=1}^{|\mathcal{E}_k|} \log_2 \left( 1 + \frac{P_k}{M_{T}} \frac{\mathcal{E}_k(e)}{N_0 W} \right),
\label{eq:mimocapacity}
\end{equation}
with the vector of channel eigenvalues per user $\mathcal{E}_{k}$, and transmission power $P_k$. The resource share $\mu_k \in (0,1]$ is the normalized (unit-less) representation of the share of time.

The transmission power is a function of the target rate, depending on the number of transmit and receive antennas. For the following configurations, \eqref{eq:mimocapacity} reduces to:

1x2 SIMO:
\begin{equation}
\begin{aligned}
& \quad P_k(R_k) = \frac{1}{\epsilon_1} \left( 2^{(\frac{R_k}{W \mu_k})} - 1 \right)
\end{aligned}
\label{eq:1x2SIMO}
\end{equation}

2x2 MIMO:
\begin{equation}
 P_k(R_k) = \frac{ - (\epsilon_1 + \epsilon_2) + \sqrt{(\epsilon_1 + \epsilon_2)^2 + 4 \epsilon_1 \epsilon_2 ( 2^{\frac{R_k}{W \mu_k}} - 1 ) }}{\epsilon_1 \epsilon_2}
\label{eq:2x2MIMO}
\end{equation}
where $\epsilon_i$ is the $i$-th member of $\mathcal{E}_k$. These equations can be extended in similar fashion to combinations with up to four transmit or receive antennas. Note that a higher number of antennas would require the general algebraic solution of polynomial equations with degree five or higher, which cannot be found in line with the Abel-Ruffini theorem. 

Like on the \ac{BS} side, the number of \ac{RF} chains used for reception at the mobile could be adapted for power saving. However, the power-saving benefit of receive \ac{AA} is much smaller than in transmit \ac{AA}, where a \ac{PA} is present in each \ac{RF} chain. Moreover, multiple receive antennas boost the useful signal power and provide a diversity gain.
Therefore, $M_{\mathrm{Rx}}$ is assumed to be fixed and set to $M_{\mathrm{Rx}} {=} 2$ in the following.

Given the transmission power \eqref{eq:1x2SIMO}, \eqref{eq:2x2MIMO}, and the power model \eqref{eq:psupply}, we derive the supply power consumption for \ac{TDMA}. With \ac{DTX} a \ac{BS} may go to sleep when all $K$ users have been served; the overall average consumption is the weighted sum of power consumed during transmission and \ac{DTX} time share~$\mu_{\mathrm{S}}$:
\begin{equation}
 P_{\mathrm{supply}}(\B{r}) = \sum_{k=1}^{K} \mu_k \left( P_{0,M_{\mathrm{T}}} + \Delta_{\mathrm{PM}} P_k(R_k) \right) + \mu_{\mathrm{S}} P_{\mathrm{S}}
\end{equation}


Consequently, we define the optimization problem with $\mu_{K+1} = \mu_{\mathrm{S}}$:
\begin{equation}
\begin{aligned}
& \underset{(\mu_1,\dots,\mu_{K+1})}{\text{minimize}} 	& P_{\mathrm{supply}}(\B{r}) 		&=  \\
&							&							& \hspace*{-25mm} \sum_{k=1}^{K} \mu_k \left( P_{0,M_{\mathrm{T}}} + \Delta_{\mathrm{PM}} P_k(R_k) \right) + \mu_{K+1} P_{\mathrm{S}} \\
& \text{subject to} 					&& \sum_{k=1}^{K+1} \mu_k 				= 1\\
&							&& \mu_k 						\geq 0\ \forall\,k \\
&							&& 0 \leq  						P_k(R_k) \leq P_{\mathrm{max}}
\end{aligned}
\label{eq:OP}
\end{equation}

The first constraint ensures that all resources are accounted for and upper bounds $\mu_k$. The second constraint encompasses the transmit power budget of the \ac{BS} and acts as a lower bound on $\mu_k$. Note that due to the block fading assumption and \ac{TDMA}, the transmission power per user $P_k$ is constant over time and equivalent to $P_t$ in~\eqref{eq:psupply}. 

This problem is convex in its cost function and constraints (proof in Appendix~\ref{appendix:proof1}). It can therefore be solved with available tools like the interior point method~\cite{book:bv0401}. As part of the \ac{RAPS} algorithm, \eqref{eq:OP} is solved once for each possible number of transmit antennas. $M_{\mathrm{T}}$ is then selected according to which solution results in lowest supply power consumption.

The solution of the first step yields an estimate for the supply power consumption, the number of transmit antennas $M_{\mathrm{T}}$, the \ac{DTX} time share $\mu_{\mathrm{S}}$ and the resource share per user $\mu_k$. If a solution for Step~1 cannot be found, Step~2 is not performed and outage occurs. We leave outage handling to a higher system layer mechanism which could be avoided by, \eg, prioritize users and reiterate with reduced system load. 

\section{Step~2: Subcarrier and Power Allocation}
\label{step2}
This section describes the second step of the \ac{RAPS} algorithm. In Step~2, the results of Step~1 are mapped back to the global problem to find the subcarrier allocation for each user $\mathcal{A}_k$, the power level per resource and spatial channel $P_{a,e}$ and the number of \ac{DTX} slots $T_{\mathrm{Sleep}}$. 

First, the real-valued resource share $\mu_k$ is mapped to the \ac{OFDMA} resource count per user $m_k \in \mathbb{N}$,
\begin{equation}
 m_k = \lceil \mu_k N T  \rceil \quad \forall k = 1,\dots,K.
\label{eq:mk}
\end{equation}
Possible rounding effects through the ceiling operation in~\eqref{eq:mk} are compensated for by adjusting the number of \ac{DTX} time slots
\begin{equation}
 T_{\mathrm{Sleep}} = \left\lfloor \frac{T N \mu_{K+1} - K}{N} \right\rfloor = \lfloor T \mu_{K+1} - K/N \rfloor.
\label{eq:tsleep}
\end{equation}
The remaining time slots are available for transmission,
\begin{equation}
 T_{\mathrm{Active}} = T - T_{\mathrm{Sleep}}.
\end{equation}
The remaining unassigned resources
\begin{equation}
 m_{\mathrm{rem}} = N T - \sum_{k=1}^{K} m_k - N T_{\mathrm{Sleep}}
\end{equation}
are assigned to $m_k$ in a round-robin fashion. After this allocation, it holds that $m_k = |\mathcal{A}_k|$.

Next, the number of assigned resource units per user $m_k$ is equally subdivided into the number of resources per user and time slot $m_{k, t}$ with
\begin{equation}
 m_{k, t} = \left \lfloor \frac{m_k}{\sum_{l=1}^{K} m_l} N \right\rfloor.
\end{equation}
The remaining unassigned resources 
\begin{equation}
 m_{t,\mathrm{rem}} = N - \sum_{k=1}^{K} m_{k,t}
\end{equation}
are allocated to different $m_{k,t}$ in a round-robin fashion.

Time slots considered for \ac{DTX} are assigned statically, starting from the back of the frame. Note that the dynamic selection of \ac{DTX} slots would create additional opportunities for capacity gains or power savings as they could be assigned to time slots with poor channel states, \eg\ time slots experiencing deep fades.

A corner case exists when $T N \mu_{K+1} < K$ and \eqref{eq:tsleep} becomes negative. This occurs when the \ac{DTX} share $\mu_{K+1} = \mu_{\mathrm{S}}$ is very small and thus traffic load is high. Due to the high traffic load, it is possible that the target rates cannot be fulfilled within the transmit power constraint, leading to outage for at least one user. Accordingly, if $T N \mu_{K+1} < K$, we set $T_{\mathrm{Active}} = T$ and the resource mapping strategy in \eqref{eq:mk} is adapted such that 
\begin{equation}
 m_k = \lfloor \mu_k N T \rfloor
\label{mkhighload}
\end{equation}
which guarantees that $m_k < NT$. The remaining resources are allocated to users as outlined above.

A subcarrier allocation algorithm from Kivanc \emph{et al.}~\cite{kll0301} is adopted, which has been shown to work effectively with low complexity. The general idea of the algorithm is as follows: First assign each subcarrier to the user with the best channel. Then start trading subcarriers from users with too many subcarriers to users with too few subcarriers based on a nearest-neighbor evaluation of the channel state. The algorithm is outlined in Algorithm~\ref{alg1} and applied to each time slot~$t$ consecutively. 

\begin{algorithm}
\caption{Adapted \ac{RCG} algorithm performed on each time slot~$t$. As compared to~\cite{kll0301} the cost parameter $h_{n,t,k}$ has been adapted and the absolute value has been added to the search of the nearest neighbor. $\mathcal{A}_{k,t}$ holds the set of subcarriers assigned to user~$k$.}
\label{alg1}
\begin{algorithmic}[1]
\ENSURE $m_{k,t}$ is the target number of subcarriers allocated to each user $k$, $h_{n,t,k} = |\overline{\B{H}_{n,t,k}}|^2$ and $\mathcal{A}_{k,t} \leftarrow\{\}$ for $k=1,\ldots,K$.
\FORALL{subcarriers $n$}
  \STATE $k^* \leftarrow \displaystyle\argmax_{1\leq k \leq K} h_{n,t,k}$
  \STATE $\mathcal{A}_{k^*,t} \leftarrow \mathcal{A}_{k^*,t} \cup \{n\}$
\ENDFOR
\FORALL{users $k$ such that $|\mathcal{A}_{k,t}| > m_{k,t}$}
  \WHILE{$|\mathcal{A}_{k,t}| > m_{k,t}$}
    \STATE $l^* \leftarrow \displaystyle\argmin_{\{l:|\mathcal{A}_{l,t}|<m_{l,t}\}} \displaystyle\min_{1\leq n \leq N} |-h_{n,t,k} + h_{n,t,l}|$
    \STATE $n^* \leftarrow \displaystyle\argmin_{1\leq n \leq N} |-h_{n,t,k} + h_{n,t,l^*}|$
    \STATE $\mathcal{A}_{k,t} \leftarrow \mathcal{A}_{k,t} / \{n^*\}, \mathcal{A}_{l^*,t} \leftarrow \mathcal{A}_{l^*,t} \cup \{n^*\}$
  \ENDWHILE
\ENDFOR
\end{algorithmic}
\end{algorithm}

At this stage, the \ac{MIMO} configuration, the number of \ac{DTX} time slots, and the subcarrier assignment are determined. Transmit powers are assigned in both spatial and time-frequency domains via an algorithm termed Inverse Water-filling (IWF). The notion of this algorithm is as follows: First, channels are sorted by quality and the water-level is initialized on the best channel, such that the bit load target is fulfilled. Then, in each iteration of the algorithm, the next best channel is added to the set of used channels, thus reducing the water-level in each step. The search is finished once the water-level is lower than the next channel metric to be added. Refer to Appendix~\ref{appendix:iwf} for the derivation. 

Let us define for each user the target bit-load
\begin{equation}
\label{eq:btarget}
 B_{\mathrm{target}, k} = R_k \tau_{\mathrm{frame}}.
\end{equation}
To fulfill the target bit-load, the following constraint must be met
\begin{equation}
 B_{\mathrm{target}, k} - \displaystyle\sum_{a=1}^{|\mathcal{A}_k|} \sum_{e=1}^{|\mathcal{E}_{a,k}|} w \tau \log_2 \left( 1 + \frac{P_{a,e} \mathcal{E}_{a,k}(e)}{N_0 w} \right) = 0,
\end{equation}
which assesses the sum capacity over a set of resources and spatial channels.

The water-level~$\nu$ can be found via an iterative search over the set~$\Omega_k$ channels that contribute a positive power
\begin{equation}
   \log_2(\nu) = \frac{1}{|\Omega_{k} |}\!\left(\frac{B_{\mathrm{target}, k}}{w \tau} - \sum_{e=1}^{|\Omega_k|} \log_2 \!\left( \frac{\tau\, \mathcal{E}_{a,k}(e)}{N_0 \log(2)} \right)\!\right).
\end{equation}
The water-level is largest on the first iteration and decreases on each iteration until it can no longer be decreased. 

\begin{figure}[p]
\centering
\includegraphics[width=0.6\textwidth]{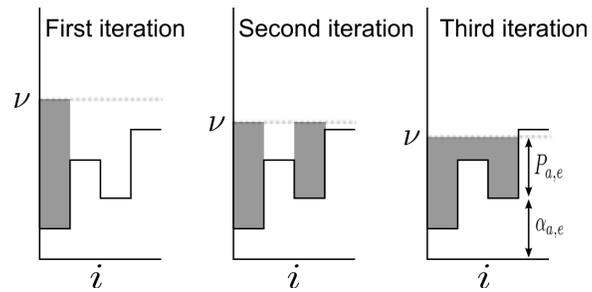}
\caption{Illustration of Inverse Water-filling (IWF) with three steps. The height of each patch~$i$ is given by $\alpha_{a,e} = N_0 w/\mathcal{E}_{a,k}(e)$. The water-level is denoted by~$\nu$. The height of the water above each patch is $P_{a,e}$. The first step sets the water-level such that the bit load target is fulfilled on the best patch. The second and third step add a patch, thus reducing the water-level. After the third step, the water-level is below the fourth patch level, thus terminating the algorithm.}
\label{fig:IWF}
\end{figure}

\begin{figure}[p]
\centering
\includegraphics[width=0.5\textwidth]{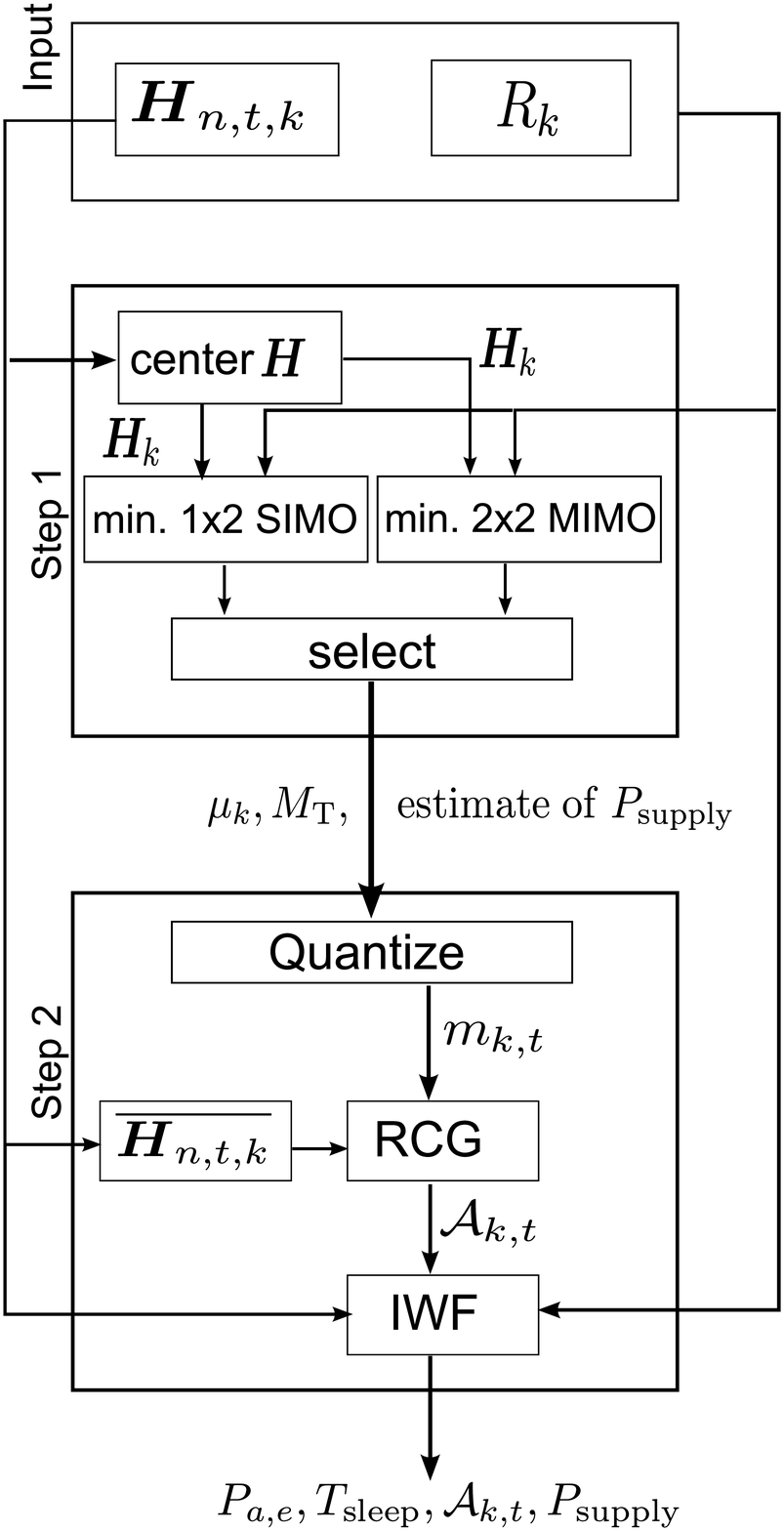}
\caption{Outline of the \ac{RAPS} algorithm.}
\label{fig:algflowchart}
\end{figure}

Using the Lagrangian method detailed in Appendix~\ref{appendix:iwf}, we arrive at a power-level per spatial channel of
\begin{equation}
 P_{a,e} = \frac{\nu w \tau }{\log(2)} - \frac{N_0 w}{\mathcal{E}_{a,k}(e)}.
\end{equation}

The inverse water-filling algorithm is illustrated for four channels in Fig.~\ref{fig:IWF}.
The outcome of Inverse Water-filling as part of \ac{RAPS} are the transmission power levels $P_{a,e}$ for each resource unit. The supply power consumption after application of \ac{RAPS} can be found by summation of transmission powers in each time slot and application of the supply power model~\eqref{eq:psupply}. The entire \ac{RAPS} algorithm is outlined in Fig.~\ref{fig:algflowchart}.

\section{Results}
\label{results}

In order to assess the performance of the \ac{RAPS} algorithm Monte Carlo simulations are conducted. The simulations are configured as follows: mobiles are uniformly distributed around the \ac{BS} on a circle with radius 250\,m and a minimum distance of 40\,m to the \ac{BS} to avoid peak \acp{SNR}. Fading is computed according to the NLOS model described in \ac{3GPP}  TR25.814~\cite{std:3gpp-plafeutra} with 8\,dB shadowing standard deviation, and the frequency-selective channel model B5 described in~\cite{std:ist-chanmod} with 3\,m/s mobile velocity. All transmit and receive antennas are assumed to be mutually uncorrelated. 
Further system parameters are listed in Table~\ref{tab:notationsummary}. 

\begin{table}[t]
\centering
\begin{tabular}{l|l|l}
      Variable 	&  & Value\\
\hline
      $K$ 				& Number of users 							& 10\\
      $N$ 				& Number of subcarriers  $\mathcal{N}$				& 50 \\
      $T$ 				& Number of time slots in set $\mathcal{T}$ 				& 10\\
      $M_{\mathrm{T}} $ 		& Number of transmit antennas 						& [1,2]\\
      $M_{\mathrm{R}}$ 			& Number of receive antennas 						& 2\\
      $P_{0,M_{\mathrm{T}}}$ 		& Circuit power consumption  				& 185\,W/260\,W\\
      $\Delta_{\mathrm{PM}} $ 		& Load-dependence factor 						& 4.7\\
      $P_{\mathrm{S}} $ 		& Power consumption in DTX 					& 150\,W\\
      $P_{\mathrm{max}} $ 		& Maximum transmission power 						& 46~dBm\\
      $\tau_{\mathrm{frame}}$/$\tau$	& Duration of frame/time slot				& 10\,ms/1\,ms\\
      $W$/$w$ 				& System/subcarrier bandwidth 						& 10\,MHz/200\,kHz\\
      $N_0$				& Noise power spectral density						& $4\times 10^{-21}$\,W/Hz\\
\end{tabular}
\caption{System Parameters}
\label{tab:notationsummary}
\end{table}

\subsection{Benchmarks}
The following transmission strategies are evaluated for comparison purposes:
\begin{itemize}
 \item The maximum \ac{BS} power consumption is obtained by constant transmission at maximum power, $P_{\mathrm{supply,max}} = P_{0,M_{\mathrm{T}}} + \Delta_{\mathrm{PM}} P_{\mathrm{max}}$. 
 \item \ac{BA}, which finds the minimum number of subcarriers that achieves the rate target. No sleep modes are utilized and all scheduled subcarriers transmit with a transmission power spectral density of $P_{\mathrm{max}}/N$. This benchmark is chosen to represent the power consumption of state-of-the-art \acp{BS} which are not capable of \ac{DTX} or \ac{AA}. 
 \item \ac{DTX} where the \ac{BS} transmits with full power $P_{\mathrm{supply,max}}$ and switches to micro sleep once the rate requirements are fulfilled.  This benchmark assesses the attainable savings when only \ac{DTX} is applied.
\end{itemize}

\subsection{Performance Analysis}
\begin{figure}
\centering
 \includegraphics[width=0.6\textwidth]{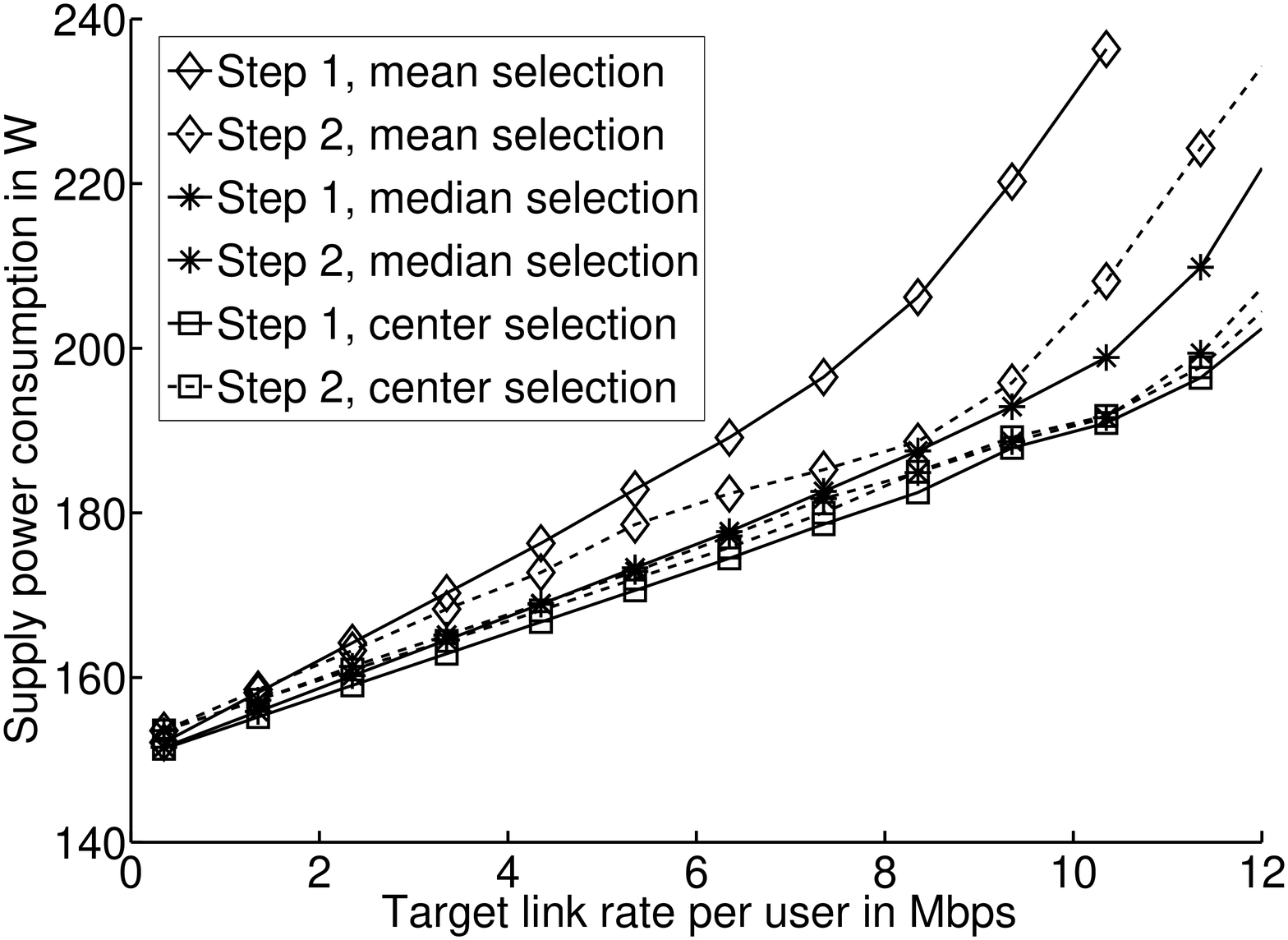}
\caption{Performance comparison of different channel state selection alternatives in Step~1.}
\label{fig:resource_selection_comparison_S1}
\end{figure}

The channel value selection in~\eqref{eq:chnCenter} serves as the channel gain for the block fading assumption of Step~1. How the channel value selection of three possible alternatives (mean, median center) affects the supply power consumption is examined in Fig.~\ref{fig:resource_selection_comparison_S1}. It can be seen that the selection of the mean channel gain results in the highest supply power consumption estimate after Step~1. While the estimate can be improved in Step~2, it is still inferior to the other alternatives. Use of the mean channel states causes Step~1 to underestimate the channel quality. Consequently, too few time slots are selected for \ac{DTX} in Step~2. 
Choosing the median provides a better estimate than the mean and after Step~2 the solution matches 'Step~2, center selection'. However, in center channel state selection the Step~1 estimate and the Step~2 solution have both the best match and the lowest supply power consumption. Therefore, as mentioned in Section~\ref{step1}, center channel state selection is chosen in the \ac{RAPS} algorithm and all further analyses.

\begin{figure}
\centering
 \includegraphics[width=0.5\textwidth]{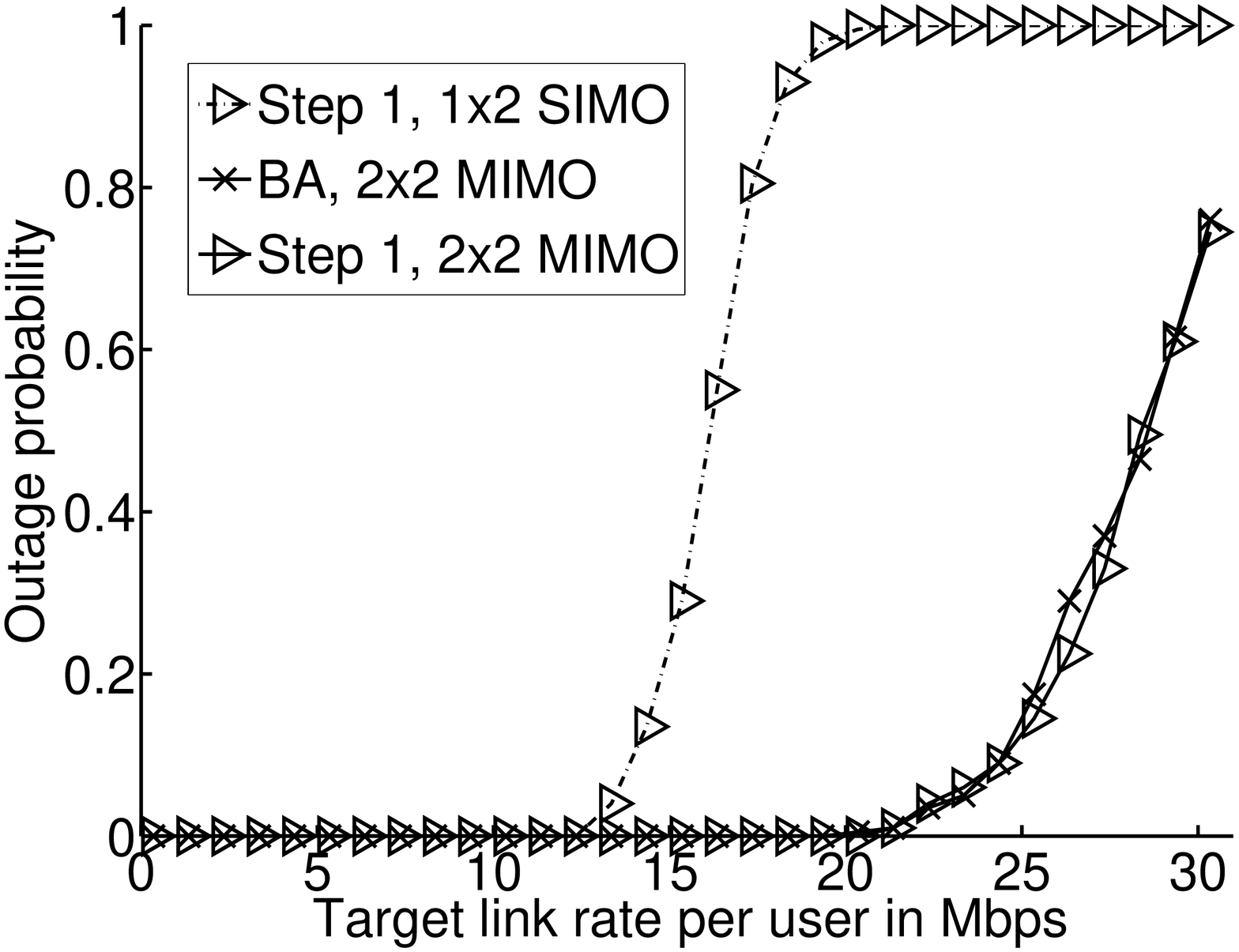}
\caption{Outage probability in Step~1 and the \ac{BA} benchmark.}
\label{fig:outage}
\end{figure}

Fig.~\ref{fig:outage} compares the outage probabilities of Step~1 with that of the \ac{BA} benchmark. Outage refers to the lack of a solution to Step~1; when the user target data rates are too high for the given channel conditions, the convex subproblem has no solution, which causes the algorithm to fail. A reduction of user target data rates based on, \eg, priorities, latency or fairness, is specifically not covered by \ac{RAPS}. One suggested alternative is to introduce admission control, in the way that some users are denied access, so that the remaining users achieve their target rates. Note that Step~2 has no effect on the outage; if a solution exists after Step~1, then Step~2 can be completed. If a solution does not exist after Step~1, then Step~2 is not performed. Fig.~\ref{fig:outage} illustrates that \ac{RAPS} selects two transmit antennas for target link rates above 14\,Mbps with high probability, as target link rates cannot be achieved with a single transmit antenna. The \ac{BA} benchmark and the Step~1 \ac{MIMO} solution have similar outage behavior.

\begin{figure}
\centering
 \includegraphics[width=0.5\textwidth]{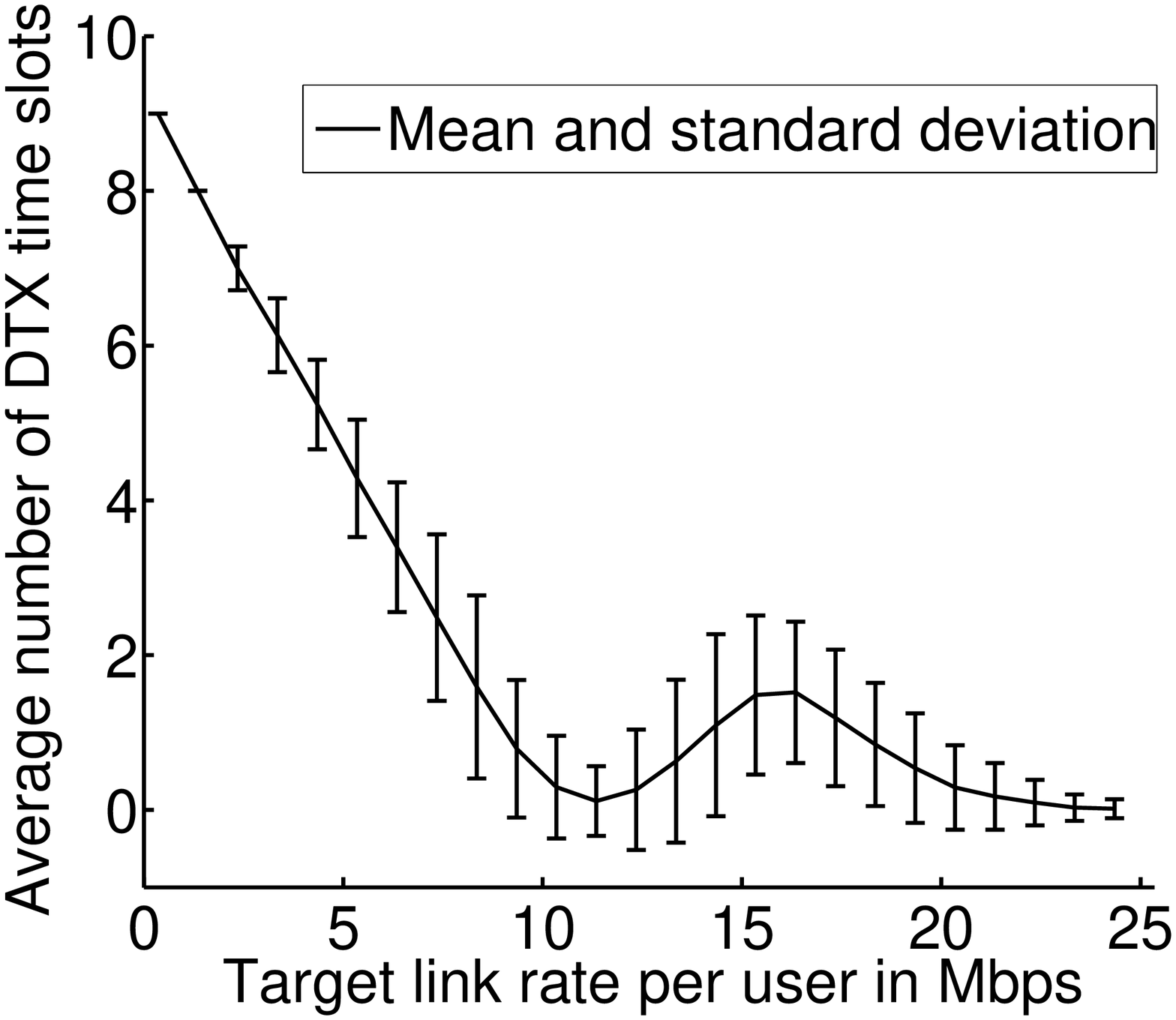}
\caption{Average number of \ac{DTX} time slots over increasing target link rates. Error bars portray the standard deviation. Total number of time slots $T = 10$.}
\label{fig:avgsleepslots}
\end{figure}

Fig.~\ref{fig:avgsleepslots} depicts the average number of \ac{DTX} time slots over increasing target link rates. The effect of \ac{AA} (\ie,~switching between \ac{SIMO} and \ac{MIMO} transmission) can clearly be seen in Fig.~\ref{fig:avgsleepslots}. For low target link rates, a large proportion of all time slots is selected for \ac{DTX}. For higher target link rates, the number of \ac{DTX} time slots must be reduced when operating in \ac{SIMO} mode. For target link rates above 12\,Mbps, which approach the \ac{SIMO} capacity (as described in the previous paragraph), the system switches to \ac{MIMO} transmission. The added \ac{MIMO} capacity allows the \ac{RAPS} scheduler to allocate more time slots to \ac{DTX}. In the medium load region (around 15~Mbps) the standard deviation is highest, indicating that here the \ac{RAPS} algorithm strongly varies the number of \ac{DTX} time slots depending on channel conditions and whether \ac{SIMO} or \ac{MIMO} transmission is selected. These variations contribute stronlgy to the additional savings provided by \ac{RAPS} over the benchmarks. Another observation from Fig.~\ref{fig:avgsleepslots} is that it is unlikely that more than two out of ten \ac{DTX} time slots are scheduled at target link rates above 8\,Mbps. This means that for a large range of target rates, no more than two \ac{DTX} time slots are required to minimize power consumption. This is an important finding for applications of \ac{RAPS} in established systems like \ac{LTE}, where the number of \ac{DTX} time slots may be limited due to constraints imposed by the standard.

\begin{figure}
\centering
 \includegraphics[width=0.6\textwidth]{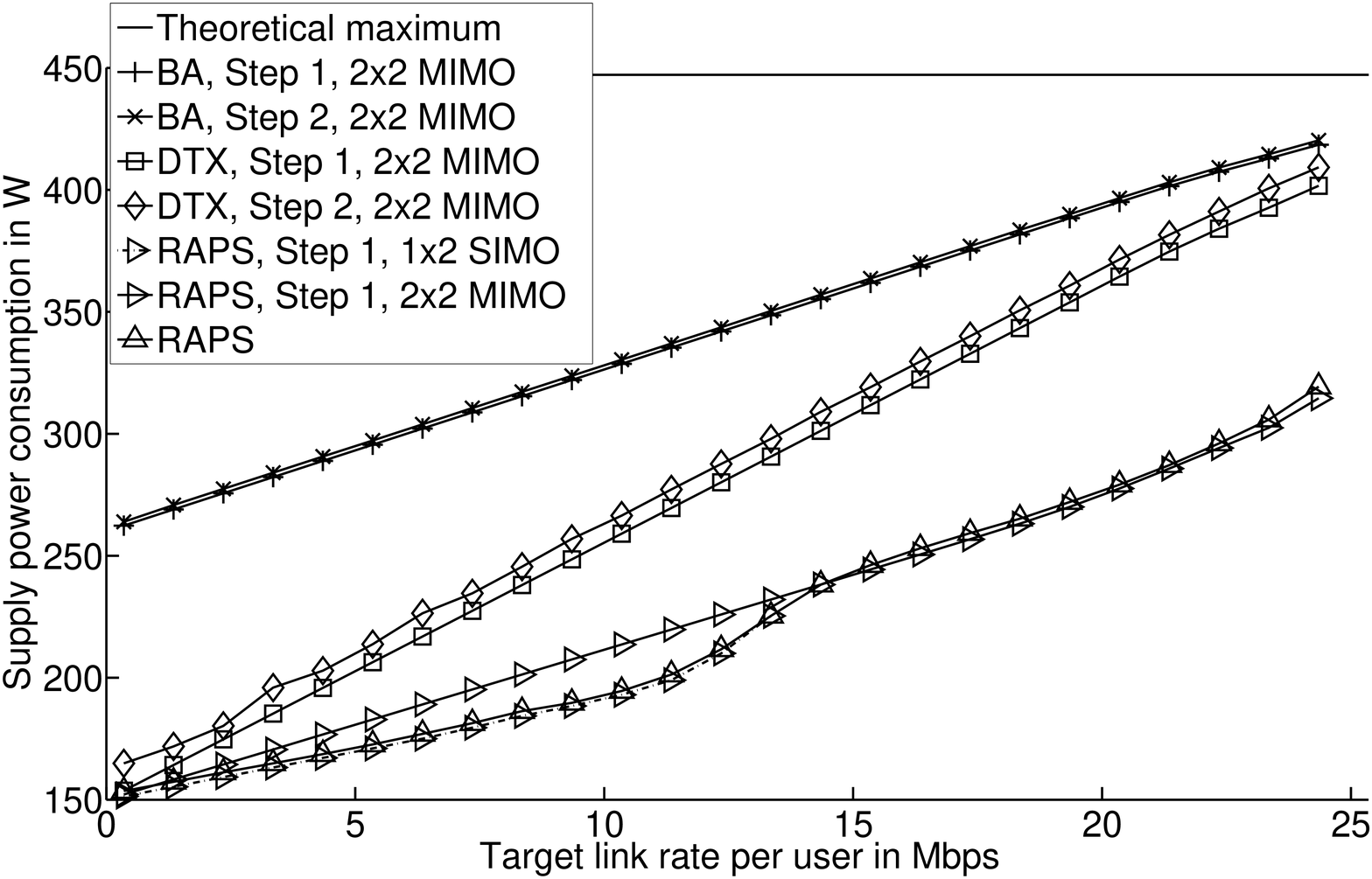}
\caption{Supply power consumption for different \ac{RRM} schemes on block fading and frequency-selective fading channels (comparison of Step~1 and Step~2) for ten users. Overlaps indicate the match between the Step~1 estimate and the Step~2 solution.}
\label{fig:fullsim_general}
\end{figure}

A comparison of the supply power consumption estimates of Step~1 and Step~2 in Fig.~\ref{fig:fullsim_general} verifies that the estimate taken in Step~1 as input for Step~2 are sufficiently accurate. Although Step~1 is greatly simplified with the assumption of block fading and its output parameters cannot be applied readily to an \ac{OFDMA} system, it precisely estimates power consumption which is the optimization cost function. The slight difference between the Step~1 estimate and power consumption after Step~2 is caused by quantization loss and resource scheduling. Note that while Step~1 supplies a good estimate of the power consumption, it is not a solution to the original \ac{OFDMA} scheduling problem, since it does not consider the frequency-selectivity of the channel and does not yield the resource and power allocation.

\begin{figure}
\centering
 \includegraphics[width=0.6\textwidth]{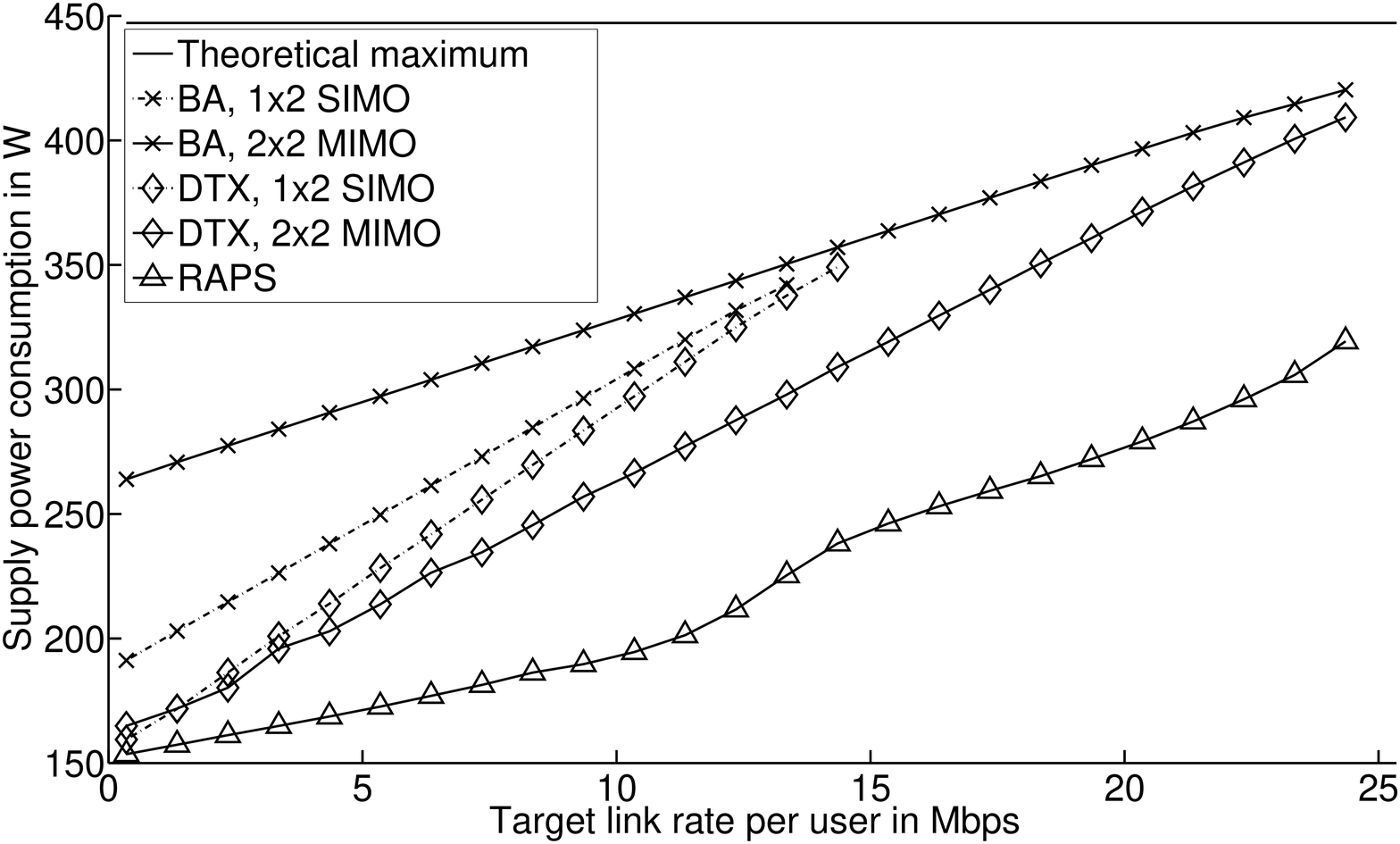}
\caption{$P_{\mathrm{supply}}$ on frequency-selective fading channels for different \ac{RRM} schemes and \ac{RAPS} for ten users. For a bandwidth-adapting \ac{BS} power consumption is always reduced by \ac{AA}, while \ac{AA} is never beneficial for a sleep mode capable \ac{BS}. For \ac{RAPS} energy consumption is reduced by \ac{AA} at low rates. In general, \ac{RAPS} achieves substantial power savings at all \ac{BS} loads.}
\label{fig:fullsim_step2}
\end{figure}

The performance of \ac{RAPS} in comparison to the benchmarks is separately analyzed in Fig.~\ref{fig:fullsim_step2}. An initial observation is that even the supply power consumption of \ac{BA} is significantly lower than the theoretical maximum for most target link rates. \ac{BA} with a single transmit antenna always consumes less power than with two antennas, as long as the rate targets with one antenna can be met. \ac{AA} is thus a valid power-saving mechanism for \ac{BA}. 

The \ac{DTX} power consumption is significantly lower than for \ac{BA}, especially at low target rates, because lower target rates allow the \ac{BS} to enter \ac{DTX} for longer periods of time. As the \ac{BS} load increases, the opportunities of the \ac{DTX} benchmark to enter sleep mode are reduced, so that \ac{DTX} power consumption approaches that of \ac{BA}. Unlike in \ac{BA}, it is never beneficial for a \ac{BS} capable of \ac{DTX} to switch operation to a single antenna, because transmitting for a longer time with a single antenna always consumes more power than a short two-antenna transmission, which allows for a longer \ac{DTX} duration. (Note that this finding may not apply to other power models.)

\ac{RAPS} reduces power consumption further than \ac{DTX} by employing \ac{AA} at low rates and \ac{PC} at high rates. Through \ac{PC}, the slope of the supply power is kept low between 15 and  20\,Mbps. At higher rates an upward trend becomes apparent, since link rates only grow logarithmically with the transmission power. In theory, when the \ac{BS} is at full load, the supply power of all energy saving mechanism will approach maximum power consumption. However, when the \ac{BS} load is very high, not all users may achieve their target rate and outage occurs (see Fig.~\ref{fig:outage}). In other words, operating the \ac{BS} with load margins controls outage and allows for large transmission power savings. Power savings of \ac{RAPS} compared to the state-of-the-art \ac{BA} range from 102.7\,W (24.5\%) to 136.9\,W (41.4\%) depending on the target link rate per user.

\begin{figure}
\centering
 \includegraphics[width=0.6\textwidth]{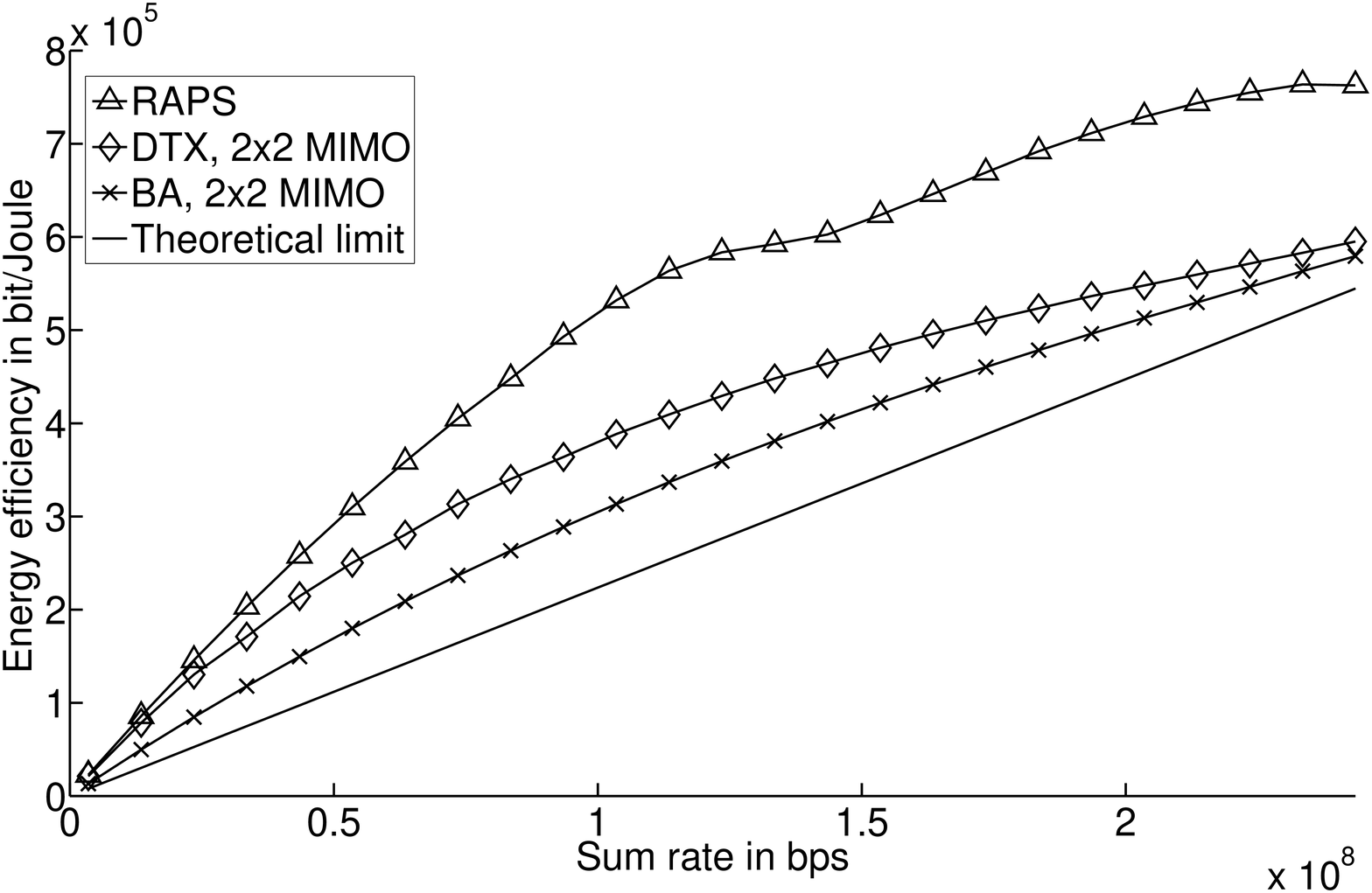}
\caption{Energy efficiency as a function of sum rate. Energy efficiency can be increased by the \ac{RAPS} algorithm. The \ac{BS} remains most efficient at peak rate.}
\label{fig:step2_EE}
\end{figure}

In addition to absolute consumption we inspect the energy efficiency of \ac{RAPS} in Fig.~\ref{fig:step2_EE}. We define energy efficiency $E = P_{\mathrm{supply}}^{-1} \sum_{k=1}^{K} R_k$ in bit/Joule. Observe that all data series are monotonically increasing. Therefore, \ac{RAPS}  does not change the paradigm that a \ac{BS} is most efficiently operated at peak rates. However, when operating below peak rates, \ac{RAPS} will always offer the most efficient operation at the requested rate.

\section{Summary and Conclusion}
\label{conclusion}

In this paper, we have examined three \ac{BS} supply power saving mechanisms, namely antenna adaptation, power control and \ac{DTX}, and evaluated their performance with a base station power consumption model. The scheduling problem of minimizing \ac{BS} supply power consumption for the downlink of multiuser \ac{MIMO}-\ac{OFDM} was defined. To solve the scheduling problem the \ac{RAPS} algorithm was proposed, which comprises two steps: first an estimate from a convex subproblem is found assuming constant channel gains on all resource units; then this estimate is applied in the second step to determine the resource and power allocation for each resource unit for a frequency-selective time-variant channel. As part of the second step, we have developed the Inverse Water-filling algorithm which finds the minimal sum power allocation for a target bit-load. Simulation results indicate that \ac{RAPS} is capable of reducing supply power consumption over all target link rates by 25\% to 40\%. Savings at low target data rates are mostly attributed to \ac{DTX} and \ac{AA}, while at high rates \ac{PC} is most effective.

While \ac{RAPS} can be applied to many types of \acp{BS}, the reported simulation results clearly depend on the used power model. Future \acp{BS} may have different power consumption characteristics. For example, smaller cells can be expected to have lower transmission power compared to circuit consumption. This work is also meant to assist hardware developments in predicting which changes will be most useful for higher-level \ac{RRM} mechanisms. Further investigations are planned into the effects of hardware switching times and hardware limitations on \ac{AA} due to the limited operating regions of \acp{PA}. Extension of \ac{RAPS} to the multi-cell interference setting is part of future work.

%
%

\appendix
\section{Proof of convexity for problem~\eqref{eq:OP}}
\label{appendix:proof1}
The partial second derivative of the \ac{MIMO} cost function in \eqref{eq:OP} with respect to $\mu_k$ reads as
\begin{equation}
\begin{aligned}
 & \frac{\partial^2 P_{\mathrm{supply},k}(\B{r})}{\partial^2 \mu_k} = \\
 & m \log^2(2) \frac{R_k}{W} \frac{ 2^{\frac{R_k}{W \mu_k}} \left( (\epsilon_1 + \epsilon_2)^2 + 2 \epsilon_1 \epsilon_2 \left( 2^{\frac{R_k}{W \mu_k}}-2 \right) \right)} {\mu_k^3 \left( (\epsilon_1 + \epsilon_2)^2 + 4 \epsilon_1 \epsilon_2 \left( 2^{\frac{R_k}{W \mu_k}}-1 \right) \right)^{\frac{3}{2}}}
\end{aligned}
\end{equation}
which is non-negative since
\begin{equation}
 2 \epsilon_1 \epsilon_2 2^{\frac{R_k}{W \mu_k}} \geq 2 \epsilon_1 \epsilon_2.
\end{equation}
Thus the cost function is convex in $\mu_k$. Convexity of the \ac{SIMO} cost function and the constraint function can be shown similarly and is omitted here for brevity.

\section{Inverse Water-filling algorithm}
\label{appendix:iwf}
Inverse Water-filling over the set of resources $\mathcal{A}_k$ and the vector of channel eigenvalues $\mathcal{E}_{a,k}$ minimizes power consumption over a set of channels while fulfilling a target bit load. 

Assuming block-diagonalization precoding, the problem reads
\begin{equation}
\begin{aligned}
& {\text{minimize}} & & \displaystyle\sum_{a=1}^{|\mathcal{A}_k|} \sum_{e=1}^{|\mathcal{E}_{a,k}|} P_{a,e}
\label{eq:iwfop}
\end{aligned}
\end{equation}
subject to
\begin{equation*}
\begin{aligned}
& B_{\mathrm{target}, k} - \displaystyle\sum_{a=1}^{|\mathcal{A}_k|} \sum_{e=1}^{|\mathcal{E}_{a,k}|} w \tau \log_2 \left( 1 + \frac{P_{a,e} \mathcal{E}_{a,k}(e)}{N_0 w} \right) = 0, \\
& P_{a,e} \geq 0 \quad \forall a,e, \\
& \displaystyle\sum_{a=1}^{|\mathcal{A}_k|} \sum_{e=1}^{|\mathcal{E}_{a,k}|} P_{a,e} \leq P_{\mathrm{max}}.
\end{aligned}
\end{equation*}

The derivative of the Lagrangian at the minimum is
\begin{equation}
 \frac{\partial \mathcal{L}}{\partial P_{a,e}} = 1 - \lambda + \beta - \frac{\nu\, w \tau\; \mathcal{E}_{a,k}(e)}{\log(2)\left( w N_0 + \mathcal{E}_{a,k}(e) P_{a,e}  \right)} = 0,
\label{eq:lagrdev}
\end{equation}
with $\lambda$ the multiplier for the first inequality constraint, $\beta$ for the second (power-limit) constraint and $\nu$ the equality constraint multiplier. The variable $\lambda$ can be omitted since it acts as a slack variable. 

From \eqref{eq:lagrdev} we arrive at a power-level per spatial channel of
\begin{equation}
 P_{a,e} = \frac{\nu w \tau }{\log(2)} - \frac{N_0 w}{\mathcal{E}_{a,k}(e)},
\end{equation}
which can be inserted into the equality constraint of \eqref{eq:iwfop} to yield the water-level $\nu$ 
\begin{equation}
   \log_2(\nu) = \frac{1}{|\Omega_k|} \left( \frac{B_{\mathrm{target}, k}}{w \tau} - \sum_{e=1}^{|\Omega_k|} \log_2 \!\left( \frac{\tau\,\mathcal{E}_{a,k}(e)}{N_0 \log(2)} \right)\right).
\end{equation}

The water-level can be found via an iterative search over the vector $\Omega_k$ of channels that contribute a positive power. Since the sum power is reduced on each iteration, the power constraint (accounted for by the multiplier~$\beta$) only needs to be tested after the search is finished.




\bibliographystyle{IEEEtran}
\bibliography{JSAC-1569687599} 


%

\vfill


\end{document}